\documentstyle[11pt]{article}
\bibliography{plain}
\title{\bf Lewenstein-Sanpera Decomposition for $2\otimes 2$ Systems } \vspace{20mm}
\author{
  S. J. Akhtarshenas$^{a,b,c}$
\thanks{E-mail:akhtarshenas@tabrizu.ac.ir}
 , M. A. Jafarizadeh$^{a,b,c}$ \thanks{E-mail:jafarizadeh@tabrizu.ac.ir}
\\
\\
$^a${\small Department of Theoretical Physics and Astrophysics,
Tabriz University, Tabriz 51664, Iran.} \\
$^b${\small Institute for Studies in Theoretical Physics and Mathematics,
 Tehran 19395-1795, Iran.} \\
$^c${\small Research Institute for Fundamental Sciences, Tabriz
51664, Iran.}} \pagebreak

\begin{document}
\maketitle \vspace{15mm}
\newpage
\begin{abstract}
As it is well known, every bipartite $2\otimes 2$ density matrix
can be obtained from Bell decomposable states via local quantum
operations and classical communications (LQCC). Using this fact,
the Lewenstein-Sanpera decomposition of an arbitrary  bipartite
$2\otimes 2$ density matrix has been obtained through LQCC action
upon  Lewenstein-Sanpera decomposition of Bell decomposable states
of $2\otimes 2$ quantum systems, where the product states
introduced by Wootters in [W. K. Wootters, Phys. Rev. Lett. {\bf
80} 2245 (1998)] form the best separable approximation ensemble
for Bell decomposable states. It is shown that in these systems
the average concurrence of the Lewenstein-Sanpera decomposition is
equal to the concurrence of these states.


{\bf PACs Index: 03.65.Ud }
\end{abstract}
\pagebreak

\vspace{7cm}

\section{Introduction}
Perhaps, quantum entanglement is the most non classical features
of quantum mechanics \cite{EPR,shcro} which has recently been
attracted much attention although it was discovered many decades
ago by Einstein and  Schr\"{O}dinger \cite{EPR,shcro}. It plays a
central role in quantum information theory and provides potential
resource for quantum communication and information processing
\cite{ben1,ben2,ben3}. Entanglement is usually arise from quantum
correlations between separated subsystems which can not be created
by local actions on each subsystems. By definition, a bipartite
mixed state $\rho$ is said to be separable if it can be expressed
as
$$
\rho=\sum_{i}w_{i}\,\rho_i^{(1)}\otimes\rho_i^{(2)},\qquad w_i\geq
0, \quad \sum_{i}w_i=1,
$$
where $\rho_i^{(1)}$ and $\rho_i^{(2)}$
denote density matrices of subsystems 1 and 2 respectively.
Otherwise the state is entangled.

The central tasks of quantum information theory is to characterize
and quantify entangled states. A first attempt in characterization
of entangled states has been made by Peres and Horodecki family
\cite{peres,horo1}. Peres showed that a necessary condition for
separability of a two partite system is that its partial
transposition be positive. Horodeckis have shown that this
condition is sufficient for separability of composite systems only
for dimensions $2\otimes 2$ and $2 \otimes 3$.

There is also an increasing attention in quantifying entanglement,
particularly for mixed states of a bipartite system, and a number
of measures have been proposed \cite{ben3,ved1,ved2,woot}. Among
them the entanglement of formation has more importance, since it
intends to quantify the resources needed to create a given
entangled state.

An interesting description of entanglement is Lewenstein-Sanpera
decomposition \cite{LS}. Lewenstein and Sanpera in \cite{LS}
showed that any two partite density matrix can be represented
optimally as a sum of a separable state and an entangled state.
They have also shown that for  2-qubit systems the decomposition
reduces to a mixture of a mixed separable state and an entangled
pure state, thus all non-separability content of the state is
concentrated in the pure entangled state. This leads to an
unambiguous measure of entanglement for any 2-qubit state as
entanglement of pure state multiplied by the weight of pure part
in the decomposition.

In the Ref. \cite{LS}, the numerical method for finding the BSA
has been reported. Also in $2\otimes2$ systems some analytical
results for special states were found in \cite{englert}. In
\cite{jaf} we have been able to  obtain an analytical expression
for L-S decomposition of Bell decomposable (BD) states. We have
also obtained the optimal decomposition for a particular class of
states obtained from BD states via some restricted LQCC actions.

In this paper using the fact that, every bipartite $2\otimes 2$
density matrix can be obtained from Bell decomposable states via
local quantum operations and classical communications
(LQCC)\cite{lind,kent,vers1,vers2}, we obtain the optimal
Lewenstein-Sanpera decomposition of an arbitrary bipart $2\otimes
2$ density matrix  through general LQCC action upon the optimal
Lewenstein-Sanpera decomposition of BD states of $2\otimes 2$
quantum systems, where the product states introduced by Wootters
in [W. K. Wootters, Phys. Rev. Lett. {\bf 80} 2245 (1998)] form
the best separable approximation ensemble for BD states. We also
show that in these systems the average concurrence of
theLewenstein-Sanpera decomposition is equal to the concurrence of
these states.

The paper is organized as follows. In section 2 we give a brief
review of Bell decomposable states together with their
separability properties. The concurrence of these states is
evaluated in section 3, via the method presented by Wootters in
\cite{woot} . In section 4 we obtain L-S decomposition of these
states. By using product states defined by Wootters in \cite{woot}
we prove that the decomposition is optimal. In section 4 we obtain
the optimal decomposition for an arbitrary $2\otimes 2$ states by
using a general LQCC action which is the main result of this
paper. The paper is ended with a brief conclusion in section 5.

\section{Bell decomposable states}
In this section we review Bell decomposable (BD) states and some
of their properties. A BD state is defined by

\begin{equation} \label{BDS}
\rho=\sum_{i=1}^{4}p_{i}\left|\psi_i\right>\left<\psi_i\right|,\quad\quad
0\leq p_i\leq 1,\quad \sum_{i=1}^{4}p_i=1.
\end{equation}
where $\left|\psi_i\right>$ are Bell states given by
\begin{eqnarray}
\label{BS1} \left|\psi_1\right>=\left|\uparrow\uparrow\right>
+\left|\downarrow\downarrow\right>), \\
\label{BS2} \left|\psi_2\right>=\left|\uparrow\uparrow\right>
-\left|\downarrow\downarrow\right>), \\
\label{BS3} \left|\psi_3\right>=\left|\uparrow\downarrow\right>
+\left|\downarrow\uparrow\right>), \\
\label{BS4} \left|\psi_4\right>=\left|\uparrow\downarrow\right>
-\left| \downarrow\uparrow\right>).
\end{eqnarray}
These states form a four simplex (tetrahedral)  with its vertices
defined by $p_1=1$, $p_2=1$, $p_3=1$ and $p_4=1$ \cite{horo2}.

A necessary condition for separability of composite quantum
systems is presented by Peres \cite{peres}. He showed that if a
state is separable then the matrix obtained from partial
transposition must be positive. Horodecki family  \cite{horo1}
have shown that Peres criterion provides sufficient condition only
for separability of mixed quantum states of dimensions $2\otimes2$
and $2\otimes3$. This implies that the state given in Eq.
(\ref{BDS}) is separable if and only if the following inequalities
are satisfying
\begin{equation}\label{ppt1}
p_i\leq \frac{1}{2},\qquad \mbox{for} \quad i=1,2,3,4.
\end{equation}
In the next sections we consider entangled states for which
$p_1\geq \frac{1}{2}$.

\section{Concurrence}
In this section we first give a brief review of concurrence of
mixed states. From the various proposed  measures of
quantification of entanglement, the entanglement of formation has
a special position which in fact intends to quantify the resources
needed to create a given entangled state \cite{ben3}. Wootters in
\cite{woot} has shown that for a 2-qubit system entanglement of
formation of a mixed state $\rho$ can be defined as
\begin{equation}
E(\rho)=H\left(\frac{1}{2}+\frac{1}{2}\sqrt{1-C^2}\right),
\end{equation}
where $H(x)=-x\ln{x}-(1-x)\ln{(1-x)}$ is binary entropy and
$C(\rho)$, called concurrence, is defined by
\begin{equation}\label{concurrence}
C(\rho)=\max\{0,\lambda_1-\lambda_2-\lambda_3-\lambda_4\},
\end{equation}
where $\lambda_i$ are the non-negative eigenvalues, with
$\lambda_1$ being the largest one, of the Hermitian matrix
$R\equiv\sqrt{\sqrt{\rho}{\tilde \rho}\sqrt{\rho}}$ and
\begin{equation}\label{rhotilde}
{\tilde \rho}
=(\sigma_y\otimes\sigma_y)\rho^{\ast}(\sigma_y\otimes\sigma_y),
\end{equation}
where $\rho^{\ast}$ is the complex conjugate of $\rho$ when it is
written in a standard basis such as
$\{\left|\uparrow\uparrow\right>,
\left|\uparrow\downarrow\right>\},\{\left|\downarrow\uparrow\right>,
\left|\downarrow\downarrow\right>\}$ and $\sigma_y$ represent
Pauli matrix in local basis $\{\left|\uparrow\right>,
\left|\downarrow\right>\}$ .

In order to obtain the concurrence of BD states we follow the
method presented by Wootters in \cite{woot}. Starting from
spectral decomposition for BD states, given in (\ref{BDS}), we
define subnormalized orthogonal eigenvectors $\left|v_i\right>$ as
\begin{equation}\label{vvector}
\left|v_i\right>=\sqrt{p_i}\left|\psi_i\right>, \qquad
\left<v_i\mid v_j\right>=p_i \delta_{ij}.
\end{equation}
Now, we can define states $\left|x_i\right>$ by
\begin{equation}\label{xvector}
\left|x_i\right>=\sum_{j}^{4}U_{ij}^{\ast}\left|v_i\right>, \qquad
\mbox{for}\quad i=1,2,3,4,
\end{equation}
such that
\begin{equation}\label{xortho}
\left<x_i\mid \tilde{x}_j\right>=(U\tau
U^T)_{ij}=\lambda_i\delta_{ij},
\end{equation}
where $\tau_{ij}=\left<v_i\mid v_j\right>$ is a symmetric but not
necessarily Hermitian matrix. To construct $\left|x_i\right>$ we
use the fact that for any symmetric matrix $\tau$ one can always
find a unitary matrix $U$ in such a way that $\lambda_i$ are real
and non-negative, that is, they are the square roots of
eigenvalues of $\tau\tau^{\ast}$ which are same as eigenvalues of
$R$. Moreover one can always find $U$ such that $\lambda_i$ appear
in decreasing order.

By using the above protocol we get for the state of $\rho$ given
in Eq. (\ref{BDS})
\begin{equation}\label{tau}
\tau=\left(\begin{array}{cccc}
-p_1 & 0 & 0 & 0 \\
 0 & p_2 & 0 & 0 \\
0 & 0 & p_3 & 0 \\
0 & 0 & 0 & -p_4
\end{array}\right).
\end{equation}
Now it is easy to evaluate $\lambda_i$ which yields
\begin{equation}\label{lambda1234}
\begin{array}{c}
\lambda_1=p_1, \quad \lambda_2=p_2, \quad \lambda_3=p_3, \quad
\lambda_4=p_4.
\end{array}
\end{equation}
Then one can evaluate the concurrence of BD states as
\begin{equation}\label{CBDS}
C=p_1-p_2-p_3-p_4=2p_1-1.
\end{equation}

Finally we introduce the unitary matrix $U$ which is going to be
used later
\begin{equation}\label{U}
U=\left(\begin{array}{cccc}
i & 0 & 0 & 0 \\
0 & 1 & 0 & 0 \\
0 & 0 & 1 & 0 \\
0 & 0 & 0 & i
\end{array}\right).
\end{equation}

\section{Lewenstein-Sanpera decomposition}
According to Lewenstein-Sanpera decomposition \cite{LS}, any
2-qubit  density matrix $\rho$ can be written as
\begin{equation}\label{LSD}
\rho=\lambda\rho_{sep}+(1-\lambda)\left|\psi\right>\left<\psi\right|,
\quad\quad \lambda\in[0,1],
\end{equation}
where $\rho_{sep}$ is a separable density matrix and
$\left|\psi\right>$ is a pure entangled state. The
Lewenstein-Sanpera decomposition of a given density matrix $\rho$
is not unique and, in general, there is a continuum set of L-S
decomposition to choose from. The optimal decomposition is,
however, unique for which $\lambda$ is maximal and
\begin{equation}\label{LSDopt}
\rho=\lambda^{(opt)}\rho_{sep}^{(opt)}
+(1-\lambda^{(opt)})|\psi^{(opt)}\left>\right<\psi^{(opt)}|\;,
\quad\quad \lambda^{(opt)}\in[0,1].
\end{equation}
 Lewenstein and Sanpera in \cite{LS} have shown that any other
  decomposition of the form
 $\rho={\tilde \lambda}{\tilde \rho}_{sep}
 +(1-{\tilde \lambda})|{\tilde \psi}\left>\right<{\tilde \psi}|$
 with ${\tilde
 \rho}\neq\rho^{(opt)}$ necessarily implies that ${\tilde
 \lambda}<\lambda^{(opt)}$ \cite{LS}.
One should notice that Eq. (\ref{LSDopt}) is the required optimal
L-S decomposition, that is, $\lambda$ is maximal and $\rho_s$ is
the best separable approximation (BSA).

Here in this section we obtain L-S decomposition for BD states.
Let us consider entangled state $\rho$ which belongs to entangled
region defined by $p_1\geq\frac{1}{2}$. We start by writing $\rho$
as a convex sum of pure state $\left|\psi_1\right>$ and separable
state $\rho_s$ as
\begin{equation}\label{LSD1}
\rho=\lambda\rho_s+(1-\lambda)\left|\psi_1\right>\left<\psi_1\right|.
\end{equation}
Expanding separable state $\rho_s$ as
$\rho_s=\sum_{i=1}^{4}p_i^{\prime}\left|\psi_i\right>\left<\psi_i\right|$
and using Eq. (\ref{BDS}) for $\rho$ we arrive at the following
results
\begin{equation}\label{pprime}
p_1^{\prime}=\frac{1}{2}, \qquad
p_i^{\prime}=\frac{p_i}{2(1-p_1)} \quad \mbox{for}\quad i=2,3,4,
\end{equation}
and
\begin{equation}\label{lambda}
\lambda=2(1-p_1).
\end{equation}
In the rest of this section we will prove that the decomposition
(\ref{LSD1}) is the optimal one. To do so we have to find a
decomposition for $\rho_s$ in terms of product states
$\left|e_\alpha,f_\alpha\right>$, i.e.
\begin{equation}
\rho_s=\sum_{\alpha}\Lambda_\alpha
\left|e_\alpha,f_\alpha\right>\left<e_\alpha,f_\alpha\right|
\end{equation}
such that  the following conditions are satisfied \cite{LS}

i) All $\Lambda_\alpha$ are maximal with respect to
$\rho_\alpha=\Lambda_\alpha
\left|e_\alpha,f_\alpha\right>\left<e_\alpha,f_\alpha\right|
+(1-\lambda)\left|\psi_1\right>\left<\psi_1\right|$ and projector
$P_\alpha=\left|e_\alpha,f_\alpha\right>\left<e_\alpha,f_\alpha\right|$.

ii) All pairs $(\Lambda_\alpha,\Lambda_\beta)$ are maximal with
respect to $\rho_{\alpha\beta}=\Lambda_\alpha
\left|e_\alpha,f_\alpha\right>\left<e_\alpha,f_\alpha\right|
+\Lambda_\beta
\left|e_\beta,f_\beta\right>\left<e_\beta,f_\beta\right|
+(1-\lambda)\left|\psi_1\right>\left<\psi_1\right|$ and the pairs
of projector $(P_\alpha,P_\beta)$.

Then according to \cite{LS} $\rho_s$ is BSA and the decomposition
given in Eq. (\ref{LSD1}) is optimal.

Lewenstein and Sanpera in \cite{LS} have shown that $\Lambda$ is
maximal with respect to $\rho$ and
$P=\left|\psi\right>\left<\psi\right|$ iff a) if
$\left|\psi\right>\not\in {\cal R}(\rho)$ then $\Lambda=0$, and b)
if $\left|\psi\right>\in {\cal R}(\rho)$ then
$\Lambda=\left<\psi\right|\rho^{-1}\left|\psi\right>^{-1}>0$. They
have also shown that a pair $(\Lambda_1,\Lambda_2)$ is maximal
with respect to $\rho$ and a pair of projectors $(P_1,P_2)$ iff:
a) if $\left|\psi_1\right>$, $\left|\psi_2\right>$ do not belong
to ${\cal R}(\rho)$ then $\Lambda_1=\Lambda_2=0$; b) if
$\left|\psi_1\right>$ does not belong, while
$\left|\psi_2\right>\in{\cal R}(\rho)$ then $\Lambda_1=0$,
$\Lambda_2=\left<\psi_2\right|\rho^{-1}\left|\psi_2\right>^{-1}$;
c) if $\left|\psi_1\right>$, $\left|\psi_2\right>\in {\cal
R}(\rho)$ and $\left<\psi_1\right|\rho^{-1}\left|\psi_2\right>=0$
then
$\Lambda_i=\left<\psi_i\right|\rho^{-1}\left|\psi_i\right>^{-1}$,
$i=1,2$; d) finally, if $\left|\psi_1\right>,
\left|\psi_2\right>\in {\cal R}(\rho)$ and
$\left<\psi_1\right|\rho^{-1}\left|\psi_2\right>\neq 0$ then

\begin{equation}\label{Lambda12}
\begin{array}{lr}
\Lambda_1= &(\left<\psi_2\right|\rho^{-1}\left|\psi_2\right>
-\mid\left<\psi_1\right|\rho^{-1}\left|\psi_2\right>\mid)/D, \\
\Lambda_2= &(\left<\psi_1\right|\rho^{-1}\left|\psi_1\right>
-\mid\left<\psi_1\right|\rho^{-1}\left|\psi_2\right>\mid)/D,
\end{array}
\end{equation}
where
$D=\left<\psi_1\right|\rho^{-1}\left|\psi_1\right>\left<\psi_2\right|
\rho^{-1}\left|\psi_2\right>
-\mid\left<\psi_1\right|\rho_{-1}\left|\psi_2\right>\mid^2$.

Now let us return to show that the decomposition given in Eq.
(\ref{LSD1}) is optimal. Wootters in \cite{woot} has shown that
any $2\otimes 2$ separable density matrix can be expanded in terms
of following product states
\begin{equation}\label{z1234-1}
\left|z_1\right>=\frac{1}{2}\left(e^{i\theta_1}\left|x_1\right>
+e^{i\theta_2}\left|x_2\right>+e^{i\theta_3}\left|x_3\right>
+e^{i\theta_4}\left|x_4\right>\right),
\end{equation}
\begin{equation}\label{z1234-1}
\left|z_2\right>=\frac{1}{2}\left(e^{i\theta_1}\left|x_1\right>
+e^{i\theta_2}\left|x_2\right>-e^{i\theta_3}\left|x_3\right>
-e^{i\theta_4}\left|x_4\right>\right),
\end{equation}
\begin{equation}\label{z1234-1}
\left|z_3\right>=\frac{1}{2}\left(e^{i\theta_1}\left|x_1\right>
-e^{i\theta_2}\left|x_2\right>+e^{i\theta_3}\left|x_3\right>
-e^{i\theta_4}\left|x_4\right>\right),
\end{equation}
\begin{equation}\label{z1234-1}
\left|z_4\right>=\frac{1}{2}\left(e^{i\theta_1}\left|x_1\right>
-e^{i\theta_2}\left|x_2\right>-e^{i\theta_3}\left|x_3\right>
+e^{i\theta_4}\left|x_4\right>\right),
\end{equation}
provided that $\lambda_1-\lambda_2-\lambda_3-\lambda_4\leq 0$.
Now, the zero concurrence is guaranteed by choosing phases
$\theta_i$, $i=1,2,3,4$ to satisfy the relation
$\sum_{j=1}e^{2i\theta_j}\lambda_j=0$.

Now using the fact that for marginal states $\rho_s$ (located at
the  boundary of separable region) the eigenvalues $\lambda_i$
satisfy constraint $\lambda_1-\lambda_2-\lambda_3-\lambda_4=0$, we
can choose the phase factors $\theta_i$ as
$\theta_2=\theta_3=\theta_4=\theta_1+\frac{\pi}{2}$. Choosing
$\theta_1=0$  we arrive at the following product ensemble for
$\rho_s$
\begin{equation}
\begin{array}{rl}
\left|z_1\right>=\frac{1}{2}
(-i\sqrt{p_1^{\prime}}\left|\psi_1\right>
-i\sqrt{p_2^{\prime}}\left|\psi_2\right>
-i\sqrt{p_3^{\prime}}\left|\psi_3\right>
-\sqrt{p_4^{\prime}}\left|\psi_4\right>),
\\
\left|z_2\right>=\frac{1}{2}
(-i\sqrt{p_1^{\prime}}\left|\psi_1\right>
-i\sqrt{p_2^{\prime}}\left|\psi_2\right>
+i\sqrt{p_3^{\prime}}\left|\psi_3\right>
+\sqrt{p_4^{\prime}}\left|\psi_4\right>),
\\
\left|z_3\right>=\frac{1}{2}
(-i\sqrt{p_1^{\prime}}\left|\psi_1\right>
+i\sqrt{p_2^{\prime}}\left|\psi_2\right>
-i\sqrt{p_3^{\prime}}\left|\psi_3\right>
+\sqrt{p_4^{\prime}}\left|\psi_4\right>),
\\
\left|z_4\right>=\frac{1}{2}
(-i\sqrt{p_1^{\prime}}\left|\psi_1\right>
+i\sqrt{p_2^{\prime}}\left|\psi_2\right>
+i\sqrt{p_3^{\prime}}\left|\psi_3\right>
-\sqrt{p_4^{\prime}}\left|\psi_4\right>),
\end{array}
\end{equation}
where $p_i^{\prime}$ are defined in Eq. (\ref{pprime}).

Let us consider the set of four product vectors
$\{\left|z_\alpha\right>\}$ and one entangled state
$\left|\psi_1\right>$. In Ref. \cite{woot} it is shown that the
ensemble $\{\left|z_\alpha\right>\}$ are linearly independent.
Evaluating Wronskian determinant of vectors $\left|\psi_1\right>$
and $\left|z_\alpha\right>$ we get $W_\alpha=\frac{1}{8}$. This
implies  that vector $\left|\psi_1\right>$ is linearly independent
with respect to all vectors $\left|z_\alpha\right>$. Also
evaluating the Wronskian of three vectors $\left|\psi_1\right>$,
$\left|z_\alpha\right>$ and $\left|z_\beta\right>$ we get
\begin{equation}\label{W}
W_{12}=W_{34}=\frac{1}{8}p_2^\prime(1-2p_2^\prime), \quad
W_{13}=W_{24}=\frac{1}{8}p_3^\prime(1-2p_3^\prime), \quad
W_{14}=W_{23}=\frac{1}{8}p_4^\prime(1-2p_4^\prime).
\end{equation}
Equations (\ref{W}) shows that in the cases that $\rho$ has full
rank three vectors $\left|z_\alpha\right>$ ,
$\left|z_\beta\right>$ and $\left|\psi_1\right>$ are linearly
independent. Now we consider the matrices
$\rho_\alpha=\Lambda_\alpha \left|
z_\alpha\right>\left<z_\alpha\right|
+(1-\lambda)\left|\psi_1\right>\left<\psi_1\right|$. Due to
independence of $\left|z_\alpha\right>$ and $\left|\psi_1\right>$
we can deduce that the range of  $\rho_\alpha$ is two dimensional.
Thus after restriction to its range and defining their dual basis
$\left|{\hat z}_\alpha\right>$ and $\left|{\hat \psi}_1\right>$,
we can expand restricted inverse $\rho_\alpha^{-1}$ as
$\rho_{\alpha}^{-1}=\Lambda_\alpha^{-1}|{\hat
z}_\alpha\left>\right<{\hat z}_\alpha| +(1-\lambda)^{-1}|{\hat
\psi}_1\left>\right<{\hat \psi}_1|$ (see appendix). Using Eq.
(\ref{A1}) it is easy to see that
$\left<z_\alpha|\rho_{\alpha}^{-1}
|z_\alpha\right>=\Lambda_\alpha^{-1}$. This shows that
$\Lambda_\alpha$ are maximal with respect to $\rho_\alpha$ and the
projector $P_\alpha$.

Similarly by considering the matrices
$\rho_{\alpha\beta}=\Lambda_\alpha
\left|z_\alpha\right>\left<z_\alpha\right| +\Lambda_\beta
\left|z_\beta\right>\left<z_\beta\right|
+(1-\lambda)\left|\psi_1\right>\left<\psi_1\right|$ and taking
into account the independence of three vectors
$\left|z_\alpha\right>$, $\left|z_\beta\right>$ and
$\left|\psi_1\right>$ we see that rang of  $\rho_{\alpha\beta}$ is
three dimensional, where after restriction to its range and
defining dual basis $\left|{\hat z}_\alpha\right>$, $\left|{\hat
z}_\beta\right>$ and $\left|{\hat \psi}_1\right>$ we can write
restricted inverse $\rho_{\alpha\beta}^{-1}$ as
$\rho_{\alpha\beta}^{-1}=\Lambda_\alpha^{-1}|{\hat
z}_\alpha\left>\right<{\hat z}_\alpha|+\Lambda_\beta^{-1}|{\hat
z}_\beta\left>\right<{\hat z}_\beta|+(1-\lambda)^{-1}|{\hat
\psi}_1\left>\right<{\hat \psi}_1|$. Then it is straightforward to
get $\left<{\hat
e}_\alpha\right|\rho_{\alpha\beta}^{-1}\left|{\hat
z}_\alpha\right>=\Lambda_\alpha^{-1}$, $\left<{\hat
z}_\beta\right|\rho_{\alpha\beta}^{-1}\left|{\hat
z}_\beta\right>=\Lambda_\beta^{-1}$ and $\left<{\hat
z}_\alpha\right|\rho_{\alpha\beta}^{-1}\left|{\hat
z}_\beta\right>=0$.

 This implies that the pairs
$(\Lambda_\alpha,\Lambda_\beta)$ are maximal with respect to
$\rho_{\alpha\beta}$ and the pairs of projectors
$(P_\alpha,P_\beta)$, hence we can conclude that the decomposition
given in Eq. (\ref{LSD1}) is optimal.

We now consider cases that $\rho$ has not full rank. Let
$p_{\alpha}=0$ for $\alpha\neq 1$. In this case Eq. (\ref{W})
shows that the pairs $\{\left|z_1\right>,\left|z_\alpha\right>\}$
and also $\{\left|z_\beta\right>,\left|z_\gamma\right>\}$ for
$\beta,\gamma\neq 1,\alpha$ are no longer independent with respect
to $\left|\psi_1\right>$. In the former case we express
$\left|\psi_1\right>$ in terms of $\left|z_1\right>$,
$\left|z_\alpha\right>$ then matrix $\rho_{1\alpha}$ can be
written in terms of two basis $\left|z_1\right>$,
$\left|z_\alpha\right>$ and after some calculations we get
$\left<z_1|\rho_{1\alpha}^{-1}|z_1\right>= \frac{\Lambda_\alpha
+2(1-\lambda)}{\Gamma_{1\alpha}}$,
$\left<z_\alpha|\rho_{1\alpha}^{-1}|z_\alpha\right>=
\frac{\Lambda_1+2(1-\lambda)}{\Gamma_{1\alpha}}$ and
$\left<z_1|\rho_{1\alpha}^{-1}|z_\alpha\right>=
\frac{-2(1-\lambda)}{\Gamma_{1\alpha}}$ where $\Gamma_{1\alpha}=
\Lambda_1\Lambda_\alpha+2(1-\lambda)(\Lambda_1+\Lambda_\alpha)$.
By using the above results together with Eqs. (\ref{Lambda12}) we
obtain the maximality of pair $(\Lambda_1,\Lambda_{\alpha})$ with
respect to $\rho_{1\alpha}$ and the pair of projectors
$(P_1,P_{\alpha})$.

Similarly for latter case we express $\left|\psi_1\right>$ in
terms of $\left|z_\beta\right>$, $\left|z_\gamma\right>$ then
matrix $\rho_{\beta\gamma}$ can be written in terms of two basis
$\left|z_\beta\right>$, $\left|z_\gamma\right>$ and we get
$\left<z_\beta|\rho_{\beta\gamma}^{-1}|z_\beta\right>=
\frac{\Lambda_\gamma +2(1-\lambda)}{\Gamma_{\beta\gamma}}$,
$\left<z_\gamma|\rho_{\beta\gamma}^{-1}|z_\gamma\right>=
\frac{\Lambda_\beta+2(1-\lambda)}{\Gamma_{\beta\gamma}}$ and
$\left<z_\beta|\rho_{\beta\gamma}^{-1}|z_\gamma\right>=
\frac{-2(1-\lambda)}{\Gamma_{\beta\gamma}}$ where
$\Gamma_{\beta\gamma}=
\Lambda_\beta\Lambda_\gamma+2(1-\lambda)(\Lambda_\beta+\Lambda_\gamma)$.
Again using the above results together with Eqs. (\ref{Lambda12})
we obtain the maximality of pairs
$(\Lambda_\beta,\Lambda_{\gamma})$ with respect to
$\rho_{\beta\gamma}$ and the pairs of projectors
$(P_\beta,P_{\gamma})$.

Finally let us consider cases that  rank $\rho$ is 2. Let
$p_{\alpha}=p_{\beta}=0$ for $\alpha,\beta\neq 1$. In this cases
we have $\left|z_\alpha\right>=\left|z_\beta\right>$  and
$\left|z_1\right>=\left|z_\gamma\right>$ for $\gamma\neq
1,\alpha,\beta$. It is now sufficient to take $\left|z_1\right>$
and $\left|z_\alpha\right>$ as product ensemble. But Eq. (\ref{W})
shows that these vectors are not independent any more, so that we
can express $\left|\psi_1\right>$ in terms of $\left|z_1\right>$
and $\left|z_\alpha\right>$, therefore, matrix $\rho_{1\alpha}$
can be written  in terms of two vectors $\left|z_1\right>$ and
$\left|z_\alpha\right>$ and we get after some calculations
$\left<z_1|\rho_{1\alpha}^{-1}|z_1\right>= \frac{\Lambda_\alpha
+2(1-\lambda)}{\Gamma_{1\alpha}}$,
$\left<z_\alpha|\rho_{1\alpha}^{-1}|z_\alpha\right>=
\frac{\Lambda_1+2(1-\lambda)}{\Gamma_{1\alpha}}$ and
$\left<z_1|\rho_{1\alpha}^{-1}|z_\alpha\right>=
\frac{-2(1-\lambda)}{\Gamma_{1\alpha}}$ where $\Gamma_{1\alpha}=
\Lambda_1\Lambda_\alpha+2(1-\lambda)(\Lambda_1+\Lambda_\alpha)$.
Using the above results together with Eqs. (\ref{Lambda12}) we
deduce the maximality of pairs $(\Lambda_1,\Lambda_{\alpha})$ with
respect to $\rho_{1\alpha}$ and the pairs of projectors
$(P_1,P_{\alpha})$.

\section{Behavior of L-S decomposition under
LQCC } In this section we study the behavior of L-S decomposition
under local quantum operations and classical communications
(LQCC). A general LQCC is defined by \cite{lind,kent}
\begin{equation}\label{lqcc}
\rho^{\prime}=\frac{(A\otimes B)\rho(A\otimes
B)^{\dag}}{Tr((A\otimes B)\rho(A\otimes B)^{\dag})},
\end{equation}
where operators $A$ and $B$ can be written as
\begin{equation}
A\otimes B=U_{A}\,f^{\mu,a,{\bf m}}\otimes U_{B}\,f^{\nu,b,{\bf
n}},
\end{equation}
where $U_{A}$ and $U_{B}$ are unitary operators acting on
subsystems $A$ and $B$, respectively and the filtration  $f$ is
defined by
\begin{equation}\label{filt}
\begin{array}{rl}
f^{\mu,a,{\bf m}}= & \mu(I_2 + a\,{\bf m}.{\bf \sigma}), \\
f^{\nu,b,{\bf n}}= & \nu(I_2 + b\,{\bf n}.{\bf \sigma}).
\end{array}
\end{equation}
As it is shown in Refs. \cite{lind,kent}, the concurrence of the
state $\rho$ transforms under LQCC of the form given in Eq.
(\ref{lqcc}) as
\begin{equation}\label{conlqcc}
C(\rho^{\prime})=\frac{\mu^2\,\nu^2(1-a^2)(1-b^2)}{Tr((A\otimes
B)\rho(A\otimes B)^{\dag})}\,C(\rho).
\end{equation}

Performing LQCC  on L-S decomposition of BD states we get

\begin{equation}\label{lqcclsd}
\rho^{\prime}= \frac{(A\otimes B)\rho(A\otimes
B)^{\dag}}{Tr((A\otimes B)\rho(A\otimes B)^{\dag})}=
\lambda^{\prime}
\rho_s^{\prime}+(1-\lambda^{\prime})\left|\psi^{\prime}\right>
\left<\psi^{\prime}\right|,
\end{equation}
with $\rho_s^{\prime}$ and $\left|\psi^{\prime}\right>$ defined as
\begin{equation}\label{lqccrhos}
\rho_s^{\prime}=\frac{(A\otimes B)\rho_s(A\otimes
B)^{\dag}}{Tr((A\otimes B)\rho_s(A\otimes B)^{\dag})},
\end{equation}
\begin{equation}
\left|\psi^{\prime}\right> =\frac{(A\otimes
B)\left|\psi_{1}\right>}{\sqrt{\left<\psi_{1}\right|(AA^{\dag}\otimes
BB^{\dag})\left|\psi_{1}\right>}},
\end{equation}
respectively, and $\lambda^{\prime}$ is
\begin{equation}\label{lqcclam}
\lambda^{\prime}=\frac{Tr((A\otimes B)\rho_s(A\otimes
B)^{\dag})}{Tr((A\otimes B)\rho(A\otimes B)^{\dag})}\,\lambda.
\end{equation}
Using Eq. (\ref{lqcclam}), we get for  the weight of entangled
part in the decomposition  (\ref{lqcclsd})
\begin{equation}
(1-\lambda^{\prime})=\frac{\left<\psi_{1}\right|(AA^{\dag}\otimes
BB^{\dag})\left|\psi_{1}\right>}{Tr((A\otimes B)\rho(A\otimes
B)^{\dag})}\,(1-\lambda).
\end{equation}
Now we can easily  evaluate the average concurrence of
$\rho^{\prime}$ in the L-S decomposition given in (\ref{lqcclsd})
\begin{equation}
(1-\lambda^{\prime})C\left(\left|\psi^{\prime}\right>\right)=
\frac{\mu^2\,\nu^2(1-a^2)(1-b^2)}{Tr((A\otimes B)\rho(A\otimes
B)^{\dag})}\,(1-\lambda)C(\left|\psi_{1}\right>),
\end{equation}
where, by comparing the above equation with Eq. (\ref{conlqcc}) we
see that $(1-\lambda)C(\left|\psi\right>)$ (the average
concurrence in the L-S decomposition) transforms  like the
concurrence under LQCC.

Now we would like to show that the decomposition given in Eq.
(\ref{lqcclsd}) is optimal. To do so, we perform  LQCC action on
matrices $\rho_{\alpha}= \Lambda_{\alpha}\left|z_{\alpha}
\right>\left<z_{\alpha}\right|+
(1-\lambda)\left|\psi_{1}\right>\left<\psi_{1}\right|$ and get
\begin{equation}
\rho_\alpha^\prime=\frac{(A\otimes B)\rho_\alpha(A\otimes
B)^{\dag}}{Tr((A\otimes B)\rho_{\alpha}(A\otimes B)^{\dag})}=
\Lambda_{\alpha}^{\prime}\left|z_{\alpha}^{\prime}
\right>\left<z_{\alpha}^{\prime}\right|+
(1-\lambda^{\prime})\left|\psi_{1}^{\prime}\right>\left<\psi_{1}^{\prime}\right|
\end{equation}
where
\begin{equation}\label{lqccpro}
\left|z_{\alpha}^{\prime}\right>= \frac{(A\otimes
B)\left|z_{\alpha}\right>}
{\sqrt{\left<z_{\alpha}\right|(AA^{\dag}\otimes
BB^{\dag})\left|z_{\alpha}\right>}},
\end{equation}
and
\begin{equation}
\Lambda_\alpha^{\prime}=\frac{\left<z_{\alpha}\right|(AA^{\dag}\otimes
BB^{\dag})\left|z_{\alpha}\right>}{Tr((A\otimes
B)\rho_{\alpha}(A\otimes B)^{\dag})}\,\Lambda_{\alpha}.
\end{equation}

Using the fact that LQCC transformations are invertible
\cite{kent,vers1,vers2}, we can evaluate
$\rho_\alpha^{\prime^{-1}}$ as
\begin{equation}
\rho_\alpha^{\prime^{-1}}=Tr((A\otimes B)\rho_{\alpha}(A\otimes
B)^{\dag})\,(A^{\dag}\otimes
B^{\dag})^{-1}\rho_\alpha^{-1}(A\otimes B)^{-1}.
\end{equation}
Using the above equation and Eq. (\ref{lqccpro}) we get
\begin{equation}\label{lamimax}
\left<z_\alpha^\prime\right|\rho_\alpha^{\prime^{-1}}
\left|z_\alpha^\prime\right>= \frac{Tr((A\otimes
B)\rho_{\alpha}(A\otimes
B)^{\dag})}{\left<z_{\alpha}\right|(AA^{\dag}\otimes
BB^{\dag})\left|z_{\alpha}\right>}
\left<z_\alpha\right|\rho_\alpha^{-1}
\left|z_\alpha\right>=\Lambda_\alpha^{\prime^{-1}}.
\end{equation}
Equation (\ref{lamimax}) shows that $\Lambda_{\alpha}^{\prime}$s
are maximal with respect to $\rho_{\alpha}^{\prime}$ and the
projector $P_{\alpha}^{\prime}$.

Matrices $\rho_{\alpha\beta}=\Lambda_{\alpha}
\left|z_{\alpha}\right>\left<z_{\alpha}\right|+\Lambda_{\beta}
\left|z_{\beta}\right>\left<z_{\beta}\right|
+(1-\lambda)\left|\psi_{1}\right>\left<\psi_{1}\right|$ transform
under LQCC as
\begin{equation}
\rho_{\alpha\beta}^{\prime}=\frac{(A\otimes
B)\rho_{\alpha,\beta}(A\otimes B)^{\dag}}{Tr((A\otimes
B)\rho_{\alpha\beta}(A\otimes B)^{\dag})}=
\Lambda_{\alpha}^{\prime}
\left|z_{\alpha}^{\prime}\right>\left<z_{\alpha}^{\prime}\right|
+\Lambda_{\beta}^{\prime}
\left|z_{\beta}^{\prime}\right>\left<z_{\beta}^{\prime}\right|
+(1-\lambda^{\prime})
\left|\psi_{1}^{\prime}\right>\left<\psi_{1}^{\prime}\right|
\end{equation}
where
\begin{equation}\label{lqccproalpha,beta}
\left|z_{\alpha,\beta}^{\prime}\right>= \frac{(A\otimes
B)\left|z_{\alpha,\beta}\right>}
{\sqrt{\left<z_{\alpha,\beta}\right|(AA^{\dag}\otimes
BB^{\dag})\left|z_{\alpha,\beta}\right>}},
\end{equation}
and
\begin{equation}
\Lambda_{\alpha,\beta}^{\prime}=
\frac{\left<z_{\alpha,\beta}\right|(AA^{\dag}\otimes
BB^{\dag})\left|z_{\alpha,\beta}\right>}{Tr((A\otimes
B)\rho_{\alpha\beta}(A\otimes B)^{\dag})}\,\Lambda_{\alpha,\beta}.
\end{equation}

We now consider cases that $\rho$ is full rank. In these cases we
have already showed that all vectors $\left|z_\alpha\right>$ ,
$\left|z_\beta\right>$ and $\left|\psi_1\right>$ are linearly
independent. Using the above results together with invertibility
of LQCC actions we arrive at the following results $$
\left<z_\alpha^\prime\right|\rho_{\alpha\beta}^{\prime^{-1}}
\left|z_\alpha^\prime\right>= \frac{Tr((A\otimes
B)\rho_{\alpha\beta}(A\otimes
B)^{\dag})}{\left<z_{\alpha}\right|(AA^{\dag}\otimes
BB^{\dag})\left|z_{\alpha}\right>}
\left<z_\alpha\right|\rho_{\alpha\beta}^{-1}
\left|z_\alpha\right>=\Lambda_\alpha^{\prime}, $$
\begin{equation}\label{lqccD12}
\left<z_\beta^\prime\left|\rho_{\alpha\beta}^{\prime^{-1}}
\right|z_\beta^\prime\right>= \frac{Tr((A\otimes
B)\rho_{\alpha\beta}(A\otimes
B)^{\dag})}{\left<z_{\beta}\right|(AA^{\dag}\otimes
BB^{\dag})\left|z_{\beta}\right>}
\left<z_\beta\right|\rho_{\alpha\beta}^{-1}
\left|z_\beta\right>=\Lambda_\beta^{\prime},
\end{equation}
$$
\left<z_\alpha^\prime\left|\rho_{\alpha\beta}^{\prime^{-1}}
\right|z_\beta^\prime\right>= \frac{Tr((A\otimes
B)\rho_{\alpha\beta}(A\otimes
B)^{\dag})}{\sqrt{\left<z_{\alpha}\right|(AA^{\dag}\otimes
BB^{\dag})\left|z_{\alpha}\right>\left<z_{\beta}\right|(AA^{\dag}\otimes
BB^{\dag})\left|z_{\beta}\right>}}
\left<z_\alpha\right|\rho_{\alpha\beta}^{-1}
\left|z_\beta\right>=0.
$$

Equations. (\ref{lqccD12}) show that the pair
$(\Lambda_{\alpha}^{\prime},\Lambda_{\beta}^{\prime})$ are maximal
with respect to $\rho_{\alpha,\beta}^{\prime}$ and the pair of
projectors $(P_{\alpha}^{\prime},P_{\beta}^{\prime})$. For other
cases that $\rho$ is not full rank we saw that there is some
dependency between three vectors $\left|z_\alpha\right>$ ,
$\left|z_\beta\right>$ and $\left|\psi_1\right>$ such that
$\left<z_\alpha\right|\rho_{\alpha\beta}^{-1})
\left|z_\beta\right>\neq 0$. This implies that in general
$\left<z_\alpha^\prime\left|\rho_{\alpha\beta}^{\prime^{-1}}
\right|z_\beta^\prime\right>\neq 0$. In this cases in \cite{jaf}
we have shown that under restricted LQCC actions for which $A=B$,
the optimality of the decomposition given in (\ref{lqcclsd}) will
be achieved.

\section{Conclusion}
We have derived Lewenstein-Sanpera decomposition for BD states and
have showed that for these states the average concurrence of the
decomposition is equal to their concurrence. It is also shown that
product states introduced by Wootters in \cite{woot} form BSA
ensemble for these states. By performing LQCC action on these
states we have been able to obtain optimal decomposition for all
$2\otimes 2$ systems. It is also shown that for these states the
average concurrence of the decomposition is equal to their
concurrence.

{\large \bf Appendix }

Let us consider the set of linearly independent vectors
$\{\left|\phi_i\right>\}$, then one can define their dual vectors
$\{\left|{\hat \phi}_i\right>\}$ such that the following relation
\begin{equation}\label{A1}
\left<{\hat \phi}_i\mid\phi_j\right>=\delta_{ij}
\end{equation}
hold. It is straightforward to show that the
$\{\left|\phi_i\right>\}$ and their dual $\{\left|{\hat
\phi}_i\right>\}$ posses the following completeness relation
\begin{equation}\label{ids}
\sum_{i}|{\hat \phi}_i\left>\right<\phi_i|=I,
\qquad\sum_{i}|\phi_i\left>\right<{\hat \phi}_i|=I.
\end{equation}
Consider an invertible     operator $M$ which is expanded in terms
of states $\left|\phi_i\right>$ as
\begin{equation}\label{M}
M=\sum_{i}a_{ij}\left|\phi_i\right>\left<\phi_j\right|
\end{equation}
Then the inverse of $M$ denoted by $M^{-1}$ can be expanded in
terms of dual bases as
\begin{equation}\label{Minverse}
M^{-1}=\sum_{i}b_{ij}|{\hat \phi}_i\left>\right<{\hat \phi}_j|
\end{equation}
where $b_{ij}=(A^{-1})_{ij}$ and $A_{ij}=a_{ij}$.


\begin{thebibliography}{99}
\bibitem{EPR}{\sc A. Einstein, B. Podolsky and Rosen, }
{\em  Phys. Rev. {\bf 47}, 777 (1935).}
\bibitem{shcro}{\sc E. Schr\"{O}dinger, }
{\em Naturwissenschaften. {\bf 23} 807 (1935).}
\bibitem{ben1}{\sc C. H. Bennett, and S. J. Wiesner,}
{\em Phys. Rev. Lett. {\bf 69}, 2881 (1992).}
\bibitem{ben2}{\sc C. H. Bennett, G. Brassard,
C. Cr\'{e}peau, R. jozsa, A Peres and W. K. Wootters,} {\em Phys.
Rev. Lett. {\bf 70}, 1895 (1993).}
\bibitem{ben3}{\sc C. H. Bennett, D. P. DiVincenzo, J. A. Smolin and W.K.
Wootters,} {\em Phys. Rev. A {\bf 54}, 3824 (1996).}
\bibitem{peres}{\sc A. Peres, }{\em Phys. Rev. Lett. {\bf 77} 1413 (1996).}
\bibitem{horo1}{\sc M. Horodecki, P. Horodecki and R. Horodecki, }
{\em Phys. Lett. A  {\bf 223} 1 (1996).}
\bibitem{ved1}{\sc V. Vedral, M. B. Plenio, M. A. Rippin and P. L. Knight,}
 {\em Phys. Rev. Lett. {\bf 78}, 2275 (1995).}
\bibitem{ved2}{\sc V. Vedral and M. B. Plenio,}
 {\em Phys. Rev. A {\bf 57}, 1619 (1998).}
\bibitem{woot}{\sc W. K. Wootters, }
{\em Phys. Rev. Lett. {\bf 80} 2245 (1998).}
\bibitem{LS}{\sc M. Lewenstein and A. Sanpera, }
 { \em Phys. Rev. Lett. {\bf 80,} 2261 (1998). }
 \bibitem{englert}{\sc B.-G Englert and N. Metwally, }
 { \em J. Mod. Opt. {\bf 47,} 2221 (2000). }
\bibitem{jaf}{\sc S. J. Akhtarshenas and M. A. Jafarizadeh, }
 { \em eprint quant-ph/0207161 {\bf }  (2002). }
\bibitem{horo2}{\sc R. Horodecki and M. Horodecki, }
{\em Phys. Rev. A {\bf 54} 1838 (1996).}
\bibitem{lind}{\sc N. Linden, S. Massar and S. Popescu, }
 {\em Phys. Rev. Lett. {\bf 81}, 3279 (1998).}
\bibitem{kent}{\sc A. Kent, N. Linden and S. Massar, }
 {\em Phys. Rev. Lett. {\bf 83}, 2656 (1999).}
\bibitem{vers1}{\sc F. Verstraete, J. Dehaene and B. DeMoor, }
 {\em Phys. Rev. A {\bf 64}, 010101 (2001).}
\bibitem{vers2}{\sc F. Verstraete, J. Dehaene and B. DeMoor, }
 {\em Phys. Rev. A {\bf 65}, 032308 (2002).}
\end{thebibliography}
\end{document}